\documentclass{article}
\usepackage{spconf,amsmath,graphicx,subcaption}
\usepackage{url, cite,booktabs}
\usepackage{amsfonts,amssymb}
\usepackage{xcolor}
\usepackage{makecell}
\usepackage[hidelinks]{hyperref}
\usepackage{svg}
\usepackage{multirow}

\vspace{-12pt}
\title{An Exploration of Task-decoupling on two-stage neural post filter for real-time personalized acoustic echo cancellation\vspace{-0.5em}}
\name{Zihan Zhang$^{1,2}$, Jiayao Sun$^1$, Xianjun Xia$^2$, Ziqian Wang$^1$, Xiaopeng Yan$^1$, Yijian Xiao$^2$, Lei Xie$^1$\vspace{-0.5em}\thanks{$^*$: Corresponding author.}}

\address{
  $^1$Audio, Speech and Language Processing Group (ASLP@NPU), School of Computer Science, \\ Northwestern Polytechnical University, Xi'an, China\\
  $^2$ByteDance, China
  \vspace{-1em}
  }
\begin{document}
%
\maketitle
\vspace{-12pt}
\begin{abstract}
Deep learning based techniques have been popularly adopted in acoustic echo cancellation (AEC).
Utilization of speaker representation has extended the frontier of AEC, thus attracting many researchers' interest in personalized acoustic echo cancellation (PAEC). Meanwhile, task-decoupling strategies are widely adopted in speech enhancement. To further explore the task-decoupling approach, we propose to use a two-stage task-decoupling post-filter (TDPF) in PAEC. Furthermore, a multi-scale local-global speaker representation is applied to improve speaker extraction in PAEC.
Experimental results indicate that the task-decoupling model can yield better performance than a single joint network. The optimal approach is to decouple the echo cancellation from noise and interference speech suppression. Based on the task-decoupling sequence, optimal training strategies for the two-stage model are explored afterwards.
\end{abstract}
\section{Introduction}
\label{sec:intro}
Recently, techniques in the area of real time communication (RTC) have developing fast to meet the high demand of online meeting, remote education and video conferences~\cite{eskimez23_interspeech}. However, the abovementioned scenarios are facing challenges derived from complicated acoustic environments thus degrading the communication quality. Speech enhancement (SE) and acoustic echo cancellation (AEC) techniques have been widely adopted to mitigate these interferences. Especially in full-duplex communication scenarios, AEC techniques become an indispensable component~\cite{zhang2022multi}. Also, presence of other interfering voices degrades the communication quality. To enhance the target voice in full-duplex scenarios, personalized AEC (PAEC) has gradually attracted researchers' attention. The latest AEC Challenge~\cite{article} also has a personalized track.


Personalized speech enhancement (PSE) has experienced rapid progress recently, which extracts the target speaker voice from the interfering and noisy signals~\cite{Wang2019,eskimez2022personalized,ju2022tea}. However, majority of the aforementioned PSE models cannot be applied to real time acoustic echo cancellation due to their high computational cost as two seperate modules. To address this limitation, some researchers introduce PAEC models~\cite{eskimez23_interspeech,zhang22t_interspeech,yu2022neuralecho} by jointly train the speaker embedding and the echo cancellation module. Most of these PAEC models utilize a multi-stage network which has shown significant success in speech enhancement~\cite{li2021icassp}. To verify the effectiveness of task-splitting approach in AEC, Braun et al.~\cite{braun2022task} demonstrate that AEC performance can be improved by performing noise suppression and echo cancellation independently. 
Although task-splitting approaches have been proved effective in AEC, how to explore an optimal task-decoupling approach for PAEC have yet to be explored due to the fact that the target speaker extractioin introduces additional complexity. 

Motivated by the multi-stage network in PSE and to explore an optimal task-decoupling approach for PAEC, we propose to use a causal two-stage task-decoupling post-filter (TDPF). The proposed approach is a hybrid approach which combines a linear filter and a neural network post-filter. Specifically, the TDPF is based on a gated convolutional F-T-LSTM neural network (GFTNN)~\cite{zhang2022multi}. To better use the target speaker information, we apply a multi-scale local-global speaker representation that combines the log-mel filterbank (FBank) and the speaker embedding from ECAPA-TDNN~\cite{desplanques2020ecapa}.


In this paper, three task-decoupling approaches are compared and analyzed, namely simple joint network, network with an extra module suppressing the residual echo at the early stage and network with an extra module suppressing interfering speaker at the later stage. Experimental results show that better performance can be achieved when echo cancellation and PSE are decoupled as two tasks at first and second stage, respectively. This approach avoids the influence of speaker embeddings on near-end speech and produces better PAEC performance. Based on the task-decoupling sequence, optimal training strategies for the two-stage model are further explored afterwards.


\vspace{-6pt}
\section{Related work}
\vspace{-3pt}
\label{sec:format}
AEC techniques based on digital signal process (DSP) have been widely applied in real-time communication (RTC) for a long time. Linear filters, including normalized least mean squares (NLMS), multi-delay block frequency domain (MDF) adaptive filters and so on, have been commonly employed for linear echo cancellation due to the ability to estimate the near-end speech or echo path. 
To address the nonlinear components, residual echo suppression (RES) algorithms are typically utilized. With the emergence of Deep Neural Networks (DNN), new possibilities for nonlinear modeling have been uncovered~\cite{zhang21ia_interspeech}, leading to active exploration of end-to-end approaches that eliminate DSP components.
West-hausen et al.~\cite{westhausen2021acoustic} incorporated the far-end signal as supplementary information to adapt the dual-signal transformation LSTM network (DTLN) for the AEC task. Zhang et al.~\cite{zhang21ia_interspeech} extended deep complex convolution recurrent network (DCCRN) with a frequency-time LSTM (F-T-LSTM) network to improve echo cancellation capabilities. Evgenii et al.~\cite{indenbom2022deep} proposed a deep learning architecture with built-in self-attention alignment, enabling the handling of unaligned inputs.
Despite the rise of end-to-end methods, hybrid approaches that combine DSP techniques with neural post-filters remain competitive and have demonstrated remarkable performance in recent AEC challenges. Peng et al.~\cite{peng2021acoustic} not only trained a gate complex convolutional recurrent neural network (GCCRN) as a post-filter but also adopted multiple filters for linear echo cancellation and time delay estimation (TDE). Zhang et al.~\cite{zhang2022multi2} designed the microphone and reference phase encoder, multi-scale time-frequency processing, and streaming axial attention to improve the performance of the RES network.

It should be noted that most of these AEC models cannot remove the potential interference speaker in near-end speech. Recently, PSE models garner researchers' attention due to their ability to remove interfering speech. Numerous PSE methods such as Voicefilter~\cite{Wang2019}, pDCCRN~\cite{eskimez2022personalized}, and TEA-PSE~\cite{ju2022tea}, achieve impressive performance in speaker extraction. 
To better improve the PSE performance, studies on the effective utilization of speaker embeddings make significant progress. He et al.~\cite{he2022local} extract speaker information on global and local levels to improve the performance of speaker extraction. Liu et al.~\cite{liu2023quantitative} show that an e-vector with a lower equal error rate (EER) does not lead to better separation, suggesting that FBank features may be more beneficial in this context. Yan et al.~\cite{yan2023npu} suggest a fusion of ResNet34-based embedding~\cite{wang2023wespeaker} and FBank. The speaker embedding from SID model is more robustness and complements with FBank. Building upon these developments, in this paper, we propose a multi-scale local-global speaker representation that incorporates multi-head cross-attention to fuse FBank and ECAPA-TDNN based speaker embedding on a global scale. A speaker encoder is used to extract speaker representation in local scale~\cite{he2022local}.

To reduce the complexity of a joint AEC-PSE model, there have been researches attempt to combine AEC with speaker extraction, known as PAEC. Zhang et al.~\cite{zhang22t_interspeech} investigated the impact of incorporating auxiliary information from far-end and near-end speaker into an AEC system. 
They developed a single-stage end-to-end network called gated temporal convolutional neural network (GTCNN).
Yu et al.~\cite{yu2022neuralecho} only focused on the near-end target speaker. They employed a two-stage network, where speaker embedding was concatenated in both stages and conducted end-to-end training. Similarly, Eskimez et al.~\cite{eskimez23_interspeech} also applied an end-to-end two-stage network. However, they constrained the focus of the first stage to echo cancellation through multi-task training. In contrast, Chen et al.~\cite{10096411} adopted a hybrid system in which the first stage of the post-filter handled denoising and echo cancellation, and the second stage focused on speaker extraction. 
While various strategies have been employed for task fusion and decoupling in denoising, echo cancellation, and speaker extraction, there has been rare study comparing these strategies within the same network architecture to determine the optimal approach. Therefore, this paper aims to identify the optimal task-decoupling approach for PAEC.
\begin{figure}[!t]
    \centering
    \includegraphics[scale=0.48]{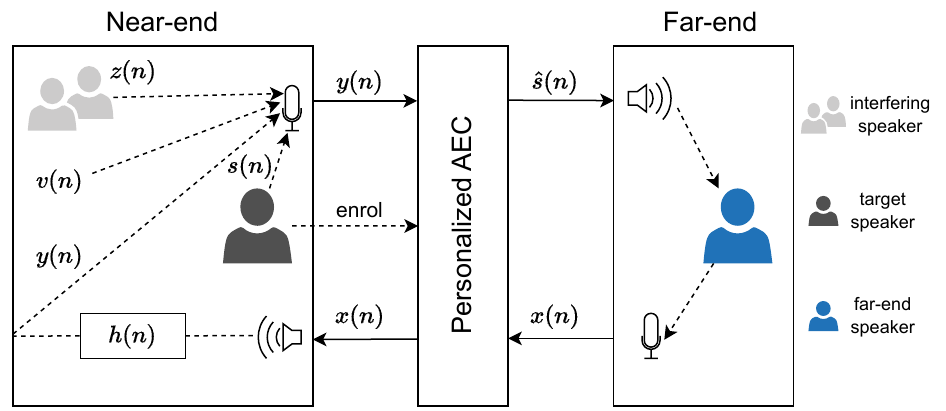} 
    \caption{The illustration of personalized AEC scenarios.} 
    \label{fig:paec_system}
    \vspace{-8pt}
\end{figure}

\section{Proposed method}
\vspace{-4pt}

\subsection{Problem formulation}
\vspace{-4pt}
We assume a duplex communication scenario as Fig.~\ref{fig:paec_system}, near-end speech can be interfered by echo, environmental noise, and interference speech. The target speech and interfering speech may exist simultaneously and have an overlap. The signal model can be expressed as
\begin{equation}
\footnotesize
\label{eq1} d(n)=s(n)+y(n)+v(n)+z(n)
\footnotesize
\end{equation}
where $d(n)$, $s(n)$, $y(n)$, $v(n)$ and $z(n)$ denote microphone signal, near-end speech signal, echo signal, additive noise and interfering speech, respectively, and $n$ is the sampling index. The PAEC system aims to extract target near-end speech $s(n)$ from $d(n)$.

\begin{figure*}[!t]
    \centering
    \includegraphics[scale=0.49]{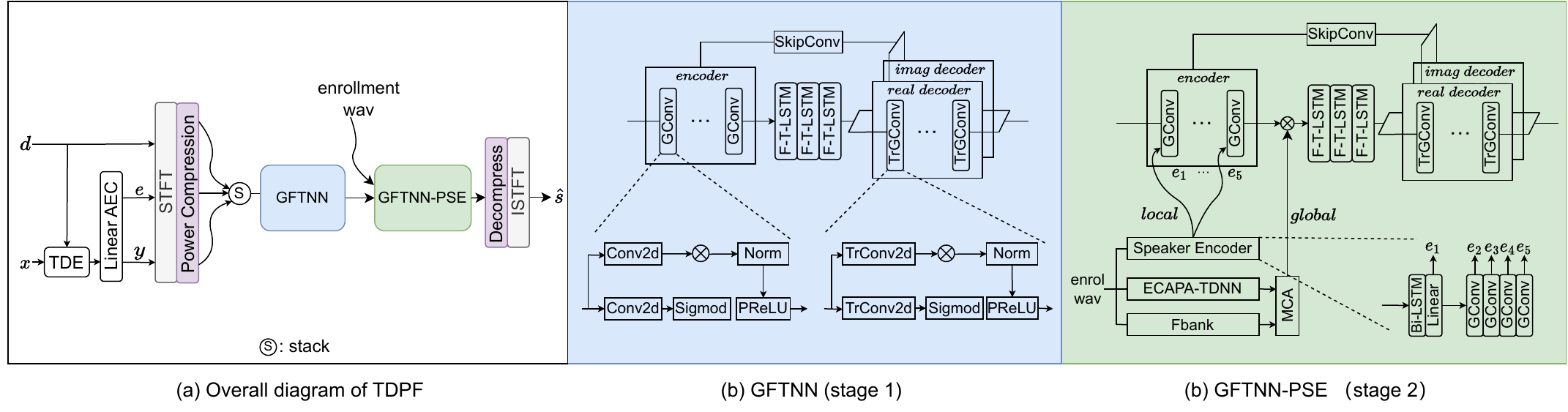}\vspace{-3pt}
    \caption{The structure of proposed two-stage network.}
    \label{fig:TDPF_model}
    \vspace{-6pt}
\end{figure*}

\vspace{-3pt}

\subsection{Two-stage post-filter}
\vspace{-3pt}
\label{sec:3.2}
For the DSP part, we employ a subband cross-correlation-based time delay estimation (TDE) algorithm to align the reference signal. A Normalized Least Mean Square (NLMS) based linear filter is used to estimate the linear echo signal $y$ and the error signal $e$.

For the post-filter, we propose a two-stage network based on GFTNN~\cite{zhang2022multi}. The gated F-T-LSTM based structure has achieved impressive results in the recent AEC challenge~\cite{article,cutler2022icassp}.
We reduce the parameters, apply the same structure in both stages and further extend it into a PAEC model by incorporating multi-scale local-global speaker representation in the second stage. Fig.~\ref{fig:TDPF_model} illustrates the model structure.
TDPF inputs the microphone signal $d$, the error signal $e$, and the linear echo $y$. 
In near-end single-talk scenarios, the reference signal should be null, but sometimes there are non-null reference signals in the training data and test data. If this situation occurs during decoding, it will seriously affect the performance of the neural network~\cite{zhang2022multi}. It's difficult for an end-to-end neural network to handle this unexpected situation. However, for the hybrid system, we can easily filter out this situation with a well-designed signal processing linear filter. Therefore, we propose a hybrid approach include linear filter and neural network post-filter.
We also apply power compression~\cite{li2021importance} with a factor of 0.5 on input signals to improve the significance of the spectrum's low-energy regions.

The first stage of the proposed model consists of five gated 2D-convolutional (Conv2d) layers in the encoder, with the input and output channel numbers remaining the same except for the first layer. To capture cross-frequency information, a bidirectional LSTM is applied on the frequency axis in the F-T-LSTM block, followed by an LSTM on the time axis to capture temporal correlations for better echo cancellation. Corresponding to the encoder, the decoder consists of five gated 2D transpose convolutional (TransConv2d) layers. Skip connections using 1D-convolutions (Conv1d) are applied between the encoder and decoder. 

The structure of the encoder, F-T-LSTM, and decoder in the second stage mirrors those of the first stage. However, we introduce an additional speaker encoder to extract local features from the enrollment wave, which are then fused into each encoder of the second stage as the local speaker representation. Moreover, multi-scale global features obtained through a multi-head cross-attention block are multiplied with the high-dimensional output of the encoder before the F-T-LSTM block. Further details regarding the multi-scale global and local speaker representations will be discussed in Section~\ref{sec:3.3}.

\vspace{-4pt}
\subsection{Multi-scale local-global speaker representation}
\vspace{-4pt}
\label{sec:3.3}
Previous studies conducted by He et al.~\cite{he2022local} and Ju et al.~\cite{ju2023tea} have revealed that incorporating a speaker encoder trained simultaneously with the PSE model, in addition to leveraging the speaker embedding provided by the speaker verification model, yields superior performance.
Inspired by this idea, we employ a bidirectional LSTM on the frequency axis to extract cross-frequency information and average all frames along the time axis to extract local features. The local features are then aligned consistently with the input dim through a fully connected layer and concatenate with each input signal frame. The speaker encoder, following the same structure as the first four layers of the second stage, functions as a parallel encoder to encoder local features and supplies to each encoder layer of TDPF.

Additionally, studies on speaker embeddings have indicated that models such as ResNet-34~\cite{wang2023wespeaker} and ECAPA-TDNN~\cite{desplanques2020ecapa}, which achieve low EER in speaker verification tasks, may discard more speaker-irrelevant information during training. However, such discarded information could be beneficial for speech separation tasks~\cite{liu2023quantitative}. 
To preserve the original information of the enrollment wave, we concatenate the temporal mean and standard deviation of 80-dimensional FBank features, resulting in a 160-dimensional embedding vector. Subsequently, we generate a 256-dimensional speaker embedding using ECAPA-TDNN~\cite{desplanques2020ecapa}. The speaker embedding is then concatenated with the FBank vector, resulting in a multi-scale speaker embedding.
To determine the most suitable speaker embedding, a multi-head cross-attention mechanism is employed, where the bottleneck features of the input audio are treated as queries (Q), the speaker embedding serves as keys (K) and values (V). The selected speaker embedding is multiplied with the input of the F-T-LSTM block, serving as the global speaker representation.

\subsection{Task-decoupling approach}
\label{sec:3.4}
In the realm of PAEC, our research aims to address several questions. First, when the number of parameters is close, is it adequate to increase the channel dimensions of the model or employ a two-stage approach? Second, for the two-stage PAEC model, which approaches yield better results: a simple joint or task-decoupling model where each stage addresses a specific task independently? Last, how should the tasks of echo cancellation, noise suppression, and target speaker extraction be allocated within the two-stage network? To provide answers, we design four comparison systems based on the network proposed in Section~\ref{sec:3.2}.
\begin{itemize}
\item \textbf{GFTNN-L} only uses the second stage of TDPF and expands the number of parameters to a level close to that of TDPF. GFTNN-L post filter receives $d$,$e$,$y$, and enrollment wav as inputs and outputs the PAEC results. Echo cancellation, noise suppression, and target speaker extraction are handled simultaneously in one large model.

\item \textbf{TDPF-1} serves as the baseline two-stage network proposed in Section~\ref{sec:3.2}. In this configuration, we do not apply task-decoupling, and the loss function is directly calculated on the final outputs.

\item \textbf{TDPF-2}, in which the first stage focuses solely on echo cancellation, accurately modeling residual echo, especially the non-linear components. The second stage performs the PSE process, removing noise and interfering speech.

\item \textbf{TDPF-3}, in which first stage simultaneously handles both echo cancellation and noise suppression, while the second stage acts as a speaker extractor, effectively separating the target speech. 

\end{itemize}

By comparing GFTNN-L with TDPF-1, TDPF-2, and TDPF-3, we can provide insights into the three aforementioned questions.

\section{Experiment setup}
\subsection{Dataset}
In our experiments, we utilize clean speech from the 4th DNS Challenge pDNS track ~\cite{dubey2022icassp} as the near-end speech, which has a sampling rate of 48k Hz and we downsample to 16k Hz for further processing. The noise signals are also sourced from the DNS-Challenge, specifically consisting of Audioset~\cite{gemmeke2017audio} and Freesound~\cite{fonseca2017freesound}, are all downsampled to 16k Hz. 

For the acoustic echo, we incorporate both real recordings and synthetic data. The real recordings are obtained from the far-end single-talk scenarios of the 4th AEC Challenge dataset. We carefully filter out entries that contain near-end speech to avoid over-suppression on target speech. The synthetic echoes are generated by convolving clean speech with the room impulse responses (RIRs) and are randomly delayed 0-0.5s. We also simulate clipping distortion or volume attenuation on 10\% of the data to make it closer to real recording.
The RIRs are generated using the HYB method~\cite{bezzam2020study}. In total, we simulate 5,000 different rooms with dimensions ranging from $3\times 3\times 3$ to $8\times 5\times 4$. For each room, we randomly select 10 positions to generate RIRs. The reverberation time (RT60) is randomly chosen from 0.2s to 1.2s. Consequently, a total of 50,000 RIRs are created. The signal-to-echo ratio (SER), which is evaluated during double-talk scenarios by Eq.(\ref{eq:ser}), randomly ranges from -15dB to 15dB. We mix the echo signal and the near-end speech based on SER. Importantly, when using synthetic echoes, we verify the speaker labels to ensure that the clean speech used for simulating the echo signal comes from a different speaker than the near-end speech.
\begin{equation}
\label{eq:ser} \mathrm{SER}=10 \log _{10}\left[\sum_{n} s^{2}(n) / \sum_{n} d^{2}(n)\right]
\end{equation}

For noise and interfering speakers, we add them to near-end speech to create mixed audio with signal-to-noise ratios (SNR) ranging from -5dB to 25dB. Interfering speakers are only added in near-end single-talk and double-talk scenarios, with a 20\% chance of no interfering speaker, a 50\% chance of one interfering speaker, and a 30\% chance of two interfering speakers. In the end, we generate a total of 400 hours of training data, with a ratio of 8:1:1 for double-talk, far-end single-talk, and near-end single-talk scenarios. We also create a validation set of 30 hours and a test set of 1500 samples. Notably, the metadata for these different datasets are completely non-overlapping, preserving the integrity and independence of each dataset.

\subsection{Performance metrics}
The effectiveness of echo cancellation can be evaluated by the echo return loss enhancement (ERLE), which provides an intuitive measurement of the performance. In the far-end single-talk (FEST) scenarios, only the echo signal is present without any near-end speech. The performance of echo cancellation can be determined by evaluating the residual of the output audio. The ERLE value is calculated using the formula provided in Eq.(\ref{eq:erle}). A higher ERLE value indicates a stronger ability for echo cancellation.
\begin{equation}
\label{eq:erle} \mathrm{ERLE}=10 \log _{10}\left[\sum_{n} d^{2}(n) / \sum_{n} \hat{s}^{2}(n)\right]
\end{equation}
In the near-end single-talk (NEST) and the double-talk (DT), we calculate wide-band perceptual evaluation of speech quality (WB-PESQ)~\cite{rix2001perceptual}. To specifically assess the performance of target speaker extraction, we calculate the scale-invariant signal-to-noise ratio (SI-SNR)~\cite{luo2018tasnet} for NEST scenarios. Additionally, we conduct AEC Mean Opinion Score (AEC MOS)~\cite{purin2021aecmos} tests on the blind test set of the 4th AEC Challenge~\cite{article}, further validating the performance of the proposed approach.

\subsection{Experimental setup}
\begin{figure*}[!t]
    \centering
    \subcaptionbox{input\label{fig:stage_out_1}}{\includegraphics[scale=0.38]{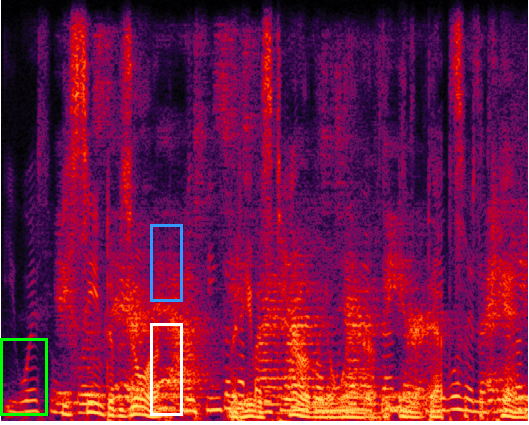}}
    \subcaptionbox{TDPF-2 stage1\label{fig:stage_out_2}}{\includegraphics[scale=0.38]{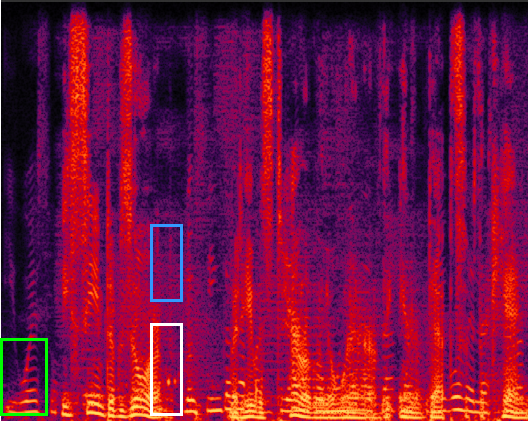}}
    \subcaptionbox{TDPF-2 stage2\label{fig:stage_out_3}}{\includegraphics[scale=0.38]{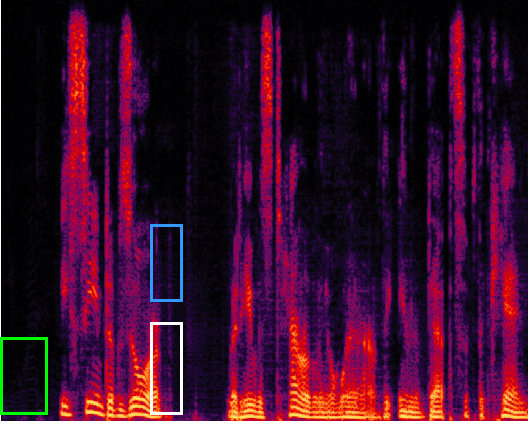}}
    \subcaptionbox{TDPF-3 stage1\label{fig:stage_out_4}}{\includegraphics[scale=0.38]{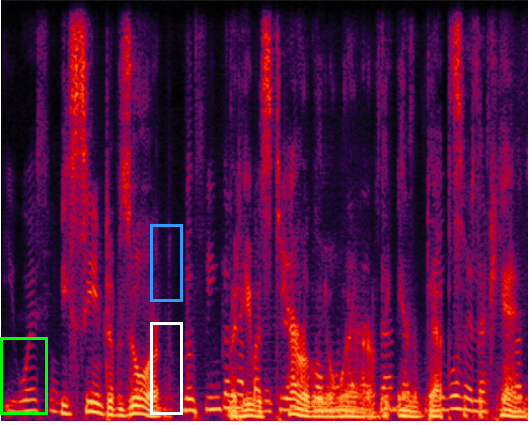}}
    \subcaptionbox{TDPF-3 stage2\label{fig:stage_out_5}}{\includegraphics[scale=0.38]{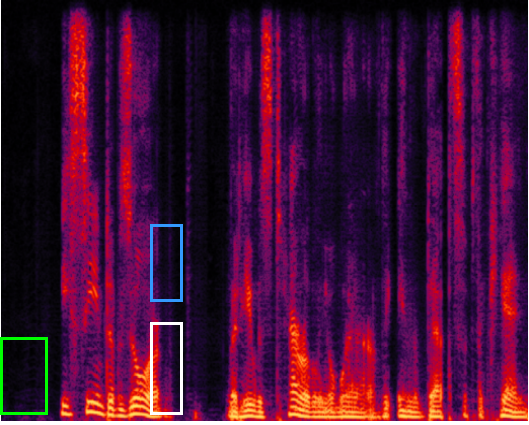}}
    \caption{Visualization of the output at each stage. The white box represents the echo component, blue box represents the noise component, green box represents interfering speech.} 
    \label{fig:stage_out}
    \vspace{-6pt}
\end{figure*}
We set the window length and hop size to 20ms and 10ms, respectively. All the Conv2d layers are configured with a kernel size of (2,3) and a stride of (1,2). The F-T-LSTM consists 128 hidden units. The Bi-LSTM in the speaker encoder has 160 hidden units, and the multi-head cross-attention is applied with 8 heads and 128 embedding dimensions. The skipconv employs a Conv1d with a kernel size of 1 and a stride of 1. Regarding GFTNN-L, which is a single-stage network, the channel number for both encoder and decoder layers is set to 160. The power-law compressed phase-aware (PLCPA) loss function is employed, defined as
\begin{equation}
\label{eq:loss1} \mathcal{L} _{GFTNN-L} = \mathcal{L}_{plcpa}(s,\hat{s}) .
\end{equation}
TDPF is a two-stage network in which the channel number of the encoder and decoder is reduced to 80. In the case of TDPF-1, task-decoupling is not applied, and the network is trained simply joint with the PLCPA loss function. In TDPF-2, the first stage aims to model the echo accurately. We initially pre-train the two stages separately, focusing on their respective tasks. Subsequently, we load the pre-trained models and perform fine-tuning. During the fine-tuning stage, the loss function is defined as 
\begin{equation}
\label{eq:loss2} \mathcal{L} _{TDPF-2} = \mathcal{L}_{plcpa}(y,\hat{s_1}) + \mathcal{L}_{plcpa}(s,\hat{s_2}) ,
\end{equation}
where $\hat{s_1}$ refers to the output of the first stage, $\hat{s_2}$ refers to the output of the second stage.

TDPF-3 follows the same structure and training strategy as TDPF-2. The difference lies in the first stage of TDPF-3, where both echo cancellation and noise suppression are performed simultaneously, and the second stage only performs speaker extraction. As a result, the loss function for TDPF-3 is defined as
\begin{equation}
\label{eq:loss3} \mathcal{L} _{TDPF-3} = \mathcal{L}_{plcpa}(s+z,\hat{s_1}) + \mathcal{L}_{plcpa}(s,\hat{s_2}) .
\end{equation}

GFTNN-AEC and GFTNN-PSE serve as the baseline for AEC and PSE tasks. The GFTNN-AEC, in fact, corresponds to the first stage of TDPF-3, with training objectives focusly on echo cancellation and noise suppression. It also serves as the pre-trained model loaded during the training of TDPF-3. 
GFTNN-PSE has the same structure as the second stage of TDPF-2 and is trained on the PSE task. To demonstrate the effectiveness of the proposed multi-scale local-global speaker representation, we replace the speaker representation component of GFTNN-PSE and train it on the same dataset. Subsequently, we evaluate its performance on the synthesized PSE test set.
These two models provide baseline performance on the subtasks, allowing us to make comparisons and assess the effectiveness of the proposed approach in the PAEC tasks.

\begin{table}[]
\centering
 \caption{PSE performance for different speaker representation in GFTNN-PSE.}
 \label{tab:1}
  \resizebox{\linewidth}{!}{
\begin{tabular}{@{}lcc@{}}
\toprule
    & WB-PESQ & SI-SNR \\ \midrule
Noisy input &  1.90  & 1.45    \\ 
ECAPA-TDNN with MCA & 2.60 & 9.46   \\
FBank and ECAPA-TDNN with MCA & 2.66 & 9.57 \\
FBank and ECAPA-TDNN with concat &   2.64   & 9.44     \\
FBank and ECAPA-TDNN with MCA + LGR& \textbf{2.70} & \textbf{9.73}   \\ 
\bottomrule
\end{tabular}
}
\vspace{-8pt}
\end{table}

\vspace{-4pt}
\section{Experiment results and analysis}
\vspace{-4pt}
In order to ensure that GFTNN-PSE model utilizes the optimal speaker representation extractor, we conducte a comparison of different speaker representations. As shown in Table~\ref{tab:1}, we have the following conclusions.

First, we observe that the fusion of ECAPA-TDNN and FBank achieve better performance compared to using only the speaker embedding from the SV model. ECAPA-TDNN exhibits superior performance in speaker verification tasks and demonstrates robustness against various forms of interference, including noise and reverberation. On the other hand, FBank retains additional speaker-irrelevant information that is discarded by ECAPA-TDNN but may useful for accurate speaker extraction.
Second, we find that multi-head cross-attention (MCA) exhibits significant advantages compared to direct concatenation. MCA allows the selection of the most appropriate speaker representation for each frame instead of applying a uniform speaker embedding across all frames. 
Last, incorporating a local-global representation (LGR) in the speaker extraction process result in improved quality. 

In summary, experimental results show that the multi-scale local-global speaker representation surpasses existing approaches and serves as a better structure for speaker extraction.

To assess the effectiveness of task-decoupling in TDPF-2 and TDPF-3, we visualize the output of each stage. In Figure~\ref{fig:stage_out}, we observe that only the echo component (highlighted in the white box) is removed in the stage 1 of TDPF-2, while the stage 2 focuses on removing noise (highlighted in the blue box) and interfering speech (highlighted in the green box). In TDPF-3, both echo and noise components are effectively removed in stage 1, and stage 2 only performs speaker extraction.
\begin{table}[]
\centering
 \caption{PAEC performance on simulate test set. ST-FE: far-end single-talk, ST-NE: near-end single-talk, DT: double-talk}
\setlength{\tabcolsep}{0.6mm}
 \label{tab:2}
  \resizebox{\linewidth}{!}{
\begin{tabular}{@{}lccccc@{}}
\toprule
    & Para. (M) & \makecell[c]{ST-FE\\(ERLE)} & \makecell[c]{DT\\(WB-PESQ)} & \makecell[c]{ST-NE\\(WB-PESQ)} & \makecell[c]{ST-NE\\(SI-SNR)} \\ \midrule
Input &  -  & 0   & 1.76 & 2.16 & 3.24   \\ 
GFTNN-AEC & 2.45 & 61.05 & 2.34 & 2.58  & 6.81  \\ 
GFTNN-PSE & 3.54 &  - & - & 2.67 & 9.98  \\ 
GFTNN-L &   7.15   &    65.83   &  2.49     & 2.85   & 11.43 \\
TDPF-1 & 6.59 & 60.9 &  2.44 &  2.75 & 9.62  \\ 
TDPF-2 & 6.59 & \textbf{66.41} & \textbf{2.55} & \textbf{2.90}  & \textbf{11.62} \\
TDPF-3 & 6.59 & 64.28 &  2.50 &  2.83  & 10.80 \\ 
\bottomrule
\vspace{-12pt}
\end{tabular}
}
\vspace{-8pt}
\end{table}
Table~\ref{tab:2} presents the PAEC performance on our simulated test set. 
It is evident that the two-stage model surpassed GFTNN-AEC and GFTNN-PSE in each individual task.
TDPF-2 also outperforms GFTNN-L, even though it has a smaller number of parameters. 
Furthermore, it can be observed that the simple joint two-stage network TSPF-1 did not achieve satisfactory results.
For PAEC, which involves multiple sub-tasks, training the two-stage model jointly without proper constraints makes it challenging to converge to an optimal state. 
Lastly, TDPF-2, which accurately models the echo in the first stage, achieves the best performance in both echo cancellation and noise suppression. 
The task of residual echo cancellation can be regarded as mapping the nonlinear echo based on the reference signal, while noise and interfering speech can be seen as interference components that can be masked from the input signal. These tasks exhibit distinct differences, and task-decoupling enables each stage to focus on different processing methods. Notably, AEC MOS metric on the blind test set of the 4th AEC Challenge in Table~\ref{tab:3} also confirms the same conclusion.

Based on these experiments, we can address the questions raised in Section~\ref{sec:3.4}: In PAEC, the two-stage model with task decoupling exhibits a higher upper bound than a large single-stage model or a simple joint two-stage model. The optimal way for task-decoupling is to separate echo cancellation from noise suppression and interference speaker removal.
\begin{table}[]
\centering
 \caption{PAEC performance on the 4th AEC challenge blind test set.}
\setlength{\tabcolsep}{0.6mm}
 \label{tab:3}
  \resizebox{\linewidth}{!}{
\begin{tabular}{@{}lcccc@{}}
\toprule
    &  \makecell[c]{ST-FE\\ECHO MOS} & \makecell[c]{DT\\Echo MOS} & \makecell[c]{DT\\Other MOS} & \makecell[c]{ST-NE\\Other MOS} \\ \midrule
Input & 2.03    & 1.71  & 3.97 & 3.92  \\ 
GFTNN-AEC & 4.60  & 4.36 & 3.95  & 4.02  \\ 
GFTNN-PSE &  -  & - & - & 4.13  \\ 
GFTNN-L &   4.73     &  4.52     & 4.03   & 4.18  \\
TDPF-1 &  4.67 & 4.36  & 3.99  & 4.12  \\ 
TDPF-2 & \textbf{4.75} & \textbf{4.55} & \textbf{4.06}  & \textbf{4.22} \\
TDPF-3 & 4.71 & 4.43  & 4.00   & 4.15 \\ 
\bottomrule
\end{tabular}
}
\vspace{-8pt}
\end{table}

We further explore the optimal training strategy under the task-decoupling approach employed in TDPF-2. Specifically, we investigate whether the first stage should be frozen and whether the second stage should be loaded during the fine-tuning of the two-stage model. As shown in Table~\ref{tab:4},
$Joint\_freeze$ refers to loading the pre-trained first stage and freezing, while does not load the second stage, joint training the second stage with the frozen first stage. $Joint$ refers to the same training approach as $Joint\_freeze$ but without freezing the first stage. $Finetune$ involves reloading both stage and fine-tuning and $finetune\_freeze$ refers to freeze the first stage during the fine-tune process.

Significant performance differences can be observed between the joint training approach and fine-tuning. Despite using the same training set for both pre-training and fine-tuning without any additional data, we still observe performance benefits from pre-training each stage on specific tasks. In practical applications, using additional training sets to pre-train each stage theoretically can lead to even higher performance improvements. 

In conclusion, for the two-stage post-filter in PAEC, the optimal structure is to have the first stage predict the echo while the second stage performs noise suppression and interfering speaker removal. The best training approach is to pre-train on each task first and then fine-tune on the PAEC training set.

\begin{table}[]
\centering
 \caption{PAEC performance for different training method on TDPF-2.}
 \label{tab:4}
  \resizebox{\linewidth}{!}{
\begin{tabular}{@{}lcccc@{}}
\toprule
     & \makecell[c]{ST-FE\\(ERLE)} & \makecell[c]{DT\\(WB-PESQ)} & \makecell[c]{ST-NE\\(WB-PESQ)} & \makecell[c]{ST-NE\\(SI-SNR)} \\ \midrule
Input &  0   & 1.76 & 2.16 & 4.24   \\ 
Joint\_freeze &  60.86 & 2.37 &  2.67 &  9.78 \\ 
Joint & 61.36  & 2.39 & 2.69 &  9.91 \\ 
Finetune\_freeze &     64.36   &  2.47     & 2.83   & 10.67 \\
Finetune &  \textbf{66.41} & \textbf{2.55} & \textbf{2.90} & \textbf{11.62}  \\ 
\bottomrule
\end{tabular}
}
\vspace{-8pt}
\end{table}

\section{Conclusions}
In this study, we proposed a two-stage post-filter in a hybrid system of linear filter and neural post-filter. We improved the speaker representation by introducing a multi-scale local-global speaker representation for more accurate speaker extraction. We explored task-decoupling on PAEC using TDPF and prove that the optimal approach involves precise echo modeling in the first stage and handling noise and interfering speech in the second stage. We also proposed a suitable training approach for the two-stage model.

\small
\bibliographystyle{IEEEbib}
\bibliography{strings,refs}

\end{document}